\documentclass[preprint,12pt]{elsarticle}

\usepackage{soul}

\usepackage{algorithm}
\usepackage{algpseudocode}
\usepackage{amsmath}
\usepackage[utf8]{inputenc}
\usepackage{subcaption}
\usepackage{graphicx}

\usepackage[utf8]{inputenc} 
\usepackage[T1]{fontenc}    
\usepackage[colorlinks, citecolor=green, linkcolor=red]{hyperref}      
\usepackage{amsfonts}       

\usepackage{tabularx, multirow, booktabs}
\usepackage{verbatim}
\usepackage{ragged2e}
\journal{arXiv}
\begin{document}

\begin{frontmatter}



\title{\textbf{Enhancing the security of image transmission in Quantum era: A Chaos-Assisted QKD Approach using entanglement}}

\author[inst1]{Raiyan Rahman\corref{cor1}}
\author[inst1]{Md Shawmoon Azad\corref{cor1}}
\author[inst1]{Mohammed Rakibul Hasan}
\author[inst1]{Syed Emad Uddin Shubha}
\author[inst1]{M.R.C. Mahdy\corref{cor2}}

\affiliation[inst1]{organization={Department of Electrical \& Computer Engineering},
            addressline={North South University}, 
            city={Dhaka},
            country={ Bangladesh}}

\cortext[cor1]{Authors contributed equally to this work}
\cortext[cor2]{Corresponding Author:}{\ead{mahdy.chowdhury@northsouth.edu}}
\begin{abstract}
The emergence of quantum computing has introduced unprecedented security challenges to conventional cryptographic systems, particularly in the domain of optical communications. This research addresses these challenges by innovatively combining quantum key distribution (QKD), specifically the E91 protocol, with logistic chaotic maps to establish a secure image transmission scheme. Our approach utilizes the unpredictability of chaotic systems alongside the robust security mechanisms inherent in quantum entanglement. The scheme is further fortified with an eavesdropping detection mechanism based on CHSH inequality, thereby enhancing its resilience against unauthorized access. Through quantitative simulations, we demonstrate the effectiveness of this scheme in encrypting images, achieving high entropy and sensitivity to the original images. The results indicate a significant improvement in encryption and decryption efficiency, showcasing the scheme's potential as a viable solution against the vulnerabilities posed by quantum computing advancements. Our research offers a novel perspective in secure optical communications, blending the principles of chaos theory with QKD to create a more robust cryptographic framework.
\end{abstract}



\end{frontmatter}


\section{Introduction}
\label{Introduction}
In today's interconnected world, the transmission of sensitive and confidential data plays a crucial role in various domains, including finance, healthcare, and national security. However, the rise of quantum computers with increasing qubit capacity poses a significant threat to the security of traditional cryptographic algorithms commonly used to protect data during transmission \cite{bernstein2017post}.
Shor's algorithm \cite{Shor_1997} had discussed a potential breach into a factorization-based cryptosystem, which has come into fruition in recent years \cite{Demonstration_of_shors_algorithm}. Cryptographic systems such as RSA\cite{rivest1978method} rely on factorization, which may face a devastating impact in the future. The system's security relies on the difficulty of factoring huge numbers, which prohibits unauthorized parties from effortlessly calculating the private key from the public key and protecting the secrecy of encrypted messages. However, this scheme is threatened at the moment since Shor's algorithm can successfully calculate factors of prime numbers with exponential speedup. According to \cite{Efficient_networks_for_quantum_factoring}, an estimated time complexity for Shor's algorithm is $\mathcal{O}(72(log(N))^{3})$, whereas in classical computer this would be approximately $\mathcal{O}(n^{3})$. An algorithm introduced by Grover in \cite{grover1996fast} also affected many of the cryptographic systems, jeopardizing the confidentiality and integrity of transmitted information.

To address this emerging challenge, researchers have turned their attention to quantum communication systems, which harness the principles of quantum mechanics to achieve secure data transmission. Quantum Key Distribution (QKD) is a promising area of research within quantum communication, focusing on establishing a secure cryptographic key between two parties over an insecure channel. QKD protocols leverage the unique properties of quantum mechanics, such as the no-cloning theorem \cite{buvzek1996quantum} and the uncertainty principle \cite{sen2014uncertainty}, to ensure that any eavesdropping attempts can be detected, thereby guaranteeing the security of the transmitted key.
In parallel, chaos-based communication systems have gained significant attention due to their inherent complexity and unpredictability \cite{riaz2008chaotic,pecora1990synchronization}. Chaotic systems exhibit sensitive dependence on initial conditions \cite{lorenz1963deterministic}, making them suitable for generating random-like signals. This randomness can be exploited for secure data encryption and decryption, as it becomes challenging for adversaries to predict the chaotic parameters necessary to decrypt the transmitted information \cite{kocarev1995general,pecora1990synchronization}. Moreover, chaotic systems offer potential benefits such as resistance to noise, interference, and attacks, making them appealing for secure communication applications \cite{bollt1997coding,wang2002chaos}.

The motivation behind our research lies in the pressing need to address the vulnerabilities in secure data transmission posed by quantum computers. As quantum computing technology advances, it is essential to develop novel cryptographic techniques and communication systems that can withstand potential attacks from these powerful machines. By combining the strengths of quantum communication and chaos-based encryption, we aim to provide a robust solution for secure data transmission. Therefore, our experimental work is to explore the feasibility and effectiveness of integrating chaotic communication systems with quantum protocols, specifically QKD, to enhance the security and robustness of optical communications. We intend to assess the performance and reliability of our proposed scheme by measuring the security achieved through end-to-end encryption. 
Our work aims to advance the field of secure data transmission by leveraging the Quantum E91 protocol and integrating chaotic systems. This innovative combination of chaos and quantum communication represents a significant advancement in the field, offering a unique and promising approach to secure data transmission. Our main contributions can be summarized as follows:
\begin{itemize}

\item Proposed a novel framework that successfully combines the E91 QKD protocol and the logistic chaotic map to securely transmit images.

\item Employed E91 as the key distribution protocol, leveraging quantum mechanics principles to enhance security during key transfer.

\item Incorporated the logistic chaotic map into the framework to provide strengthen the overall security and robustness of the communication system.

\item Evaluated the performance and reliability of the proposed scheme by measuring the security achieved through end-to-end encryption.
\end{itemize}

Our paper is organized in the following way: Section \ref{Sec:literature} reviews existing research and their developments, laying the groundwork for our proposed method. Section \ref{Sec:Preliminaries} outlines the proposed scheme's foundational concepts. Section \ref{Sec:Architecture} presents a detailed step-by-step description of our proposed architecture. In Section \ref{Sec:Implementation}, we introduce the tools employed for the implementation of the proposed work. Section \ref{Sec:results} presents the experimental findings of the study. Section \ref{Sec:Discussion} delves into a comprehensive analysis of the results, interpreting the findings and extracting key insights. Finally, Section \ref{Sec:Conclusion} concludes the paper by suggesting future research directions and highlighting our work's potential applications.

\section{Existing Research and Developments}\label{Sec:literature}
Over the past century, numerous mathematicians have made significant contributions by introducing a diverse range of chaos functions. Among these notable chaos functions are Lorentz Chaos \cite{Lorentz_original_paper}, Logistic Chaos \cite{Logistic_map_analysis}, and the Henon Map \cite{benedicks1991dynamics}. These remarkable mathematical constructs exhibit a fascinating property: they are highly sensitive to initial conditions and have the ability to generate pseudo-randomness. In the original Lorentz \cite{Lorentz_original_paper} paper, they discuss equations used for modeling convection that depend on three different parameters. Out of several key contributions of the papers, one that stood out is the sensitivity to initial conditions and the generation of pseudo-randomness. This introduced the possibility of using chaos maps in modern cryptography.
With the invention of non-periodic oscillation in the 1980s, cryptographic applications of chaos were discussed using Chua's circuit \cite{chua's_circuit}. Chua's circuit is a simple electrical circuit that generates a non-repeating oscillating waveform, which can be modeled with mathematical equations. This work explored the idea of using non-repeating pseudo-randomness as a one-time pad (OTP) for security protocols. \\
 \cite{bianco-1990} used the logistic map to generate floating point numbers that are XOR'd with plain text to create cipher text. The authors claim that the irreversible conversion from floating-point numbers to binary values prevents the recovery of the original values. However, it is noted that converting chaotic orbits to symbol sequences allows for the calculation of the initial condition with higher accuracy than the size of the symbol sets. This implies a potential vulnerability in the encryption scheme. Another concern raised was that the use of floating-point arithmetic in the method makes it machine-dependent and requires caution during software implementation. color image encryption scheme based on one-time keys and robust chaotic maps.\\
One recent image encryption scheme that utilizes chaos \cite{Jakub_Oravec} came up with the brilliant idea of using pixels, or more precisely blocks of pixels, to generate parameter values of Logistic Map (LM). Later, these parameter values are converted into a key string that is later used to decrypt the image. This dependency of the key on the image makes it much more robust and secure as a cryptography method. \cite{janani2021secure} has explored two layer quantum cryptosystem where image based initial seed values are generated following quantum block-based scrambling. \cite{liu2010color} describes a color image encryption scheme based on one-time keys and robust chaotic maps.\cite{liu2012image} introduces DNA complementary based encryption, which also utilizes chaos map. The method uses two techniques: confusing the pixels by randomly transforming nucleotides into their base pairs, and generating new keys based on the original image and common keys. This makes the initial conditions of the chaotic maps change with each encryption process, enhancing security. The method was tested on grayscale images of various sizes and found to be effective and resistant to common attacks. \\
However, all of these above-mentioned works rely on sharing keys using some form of classical channel. Peter Shor's work \cite{Shor_1997} shows how, hypothetically, quantum computers with sufficient qubits can break encryption algorithms that rely on prime factorization. Hence, it will be a risky procedure to share any data using a classical channel in the near future. Luckily for us, there exists a quantum secure transmission algorithm that utilizes the idea of quantum entanglement and superposition. Some of the earliest work on quantum cryptographic systems are attributed to Charles 
Bennett and Gilles Brassard \cite{BB84}. Their work proposes a quantum secure scheme that is based on Heisenberg's uncertainty principle. Arthur K. Ekert explored a hypothetical scheme based on Bell's theorem \cite{ekert-1991}. He uses EPR pair particles to conduct this scheme. EPR refers to Einstein, Rosen, and Podolsky's 1935 paper titled "Can Quantum Mechanical Description of Physical Reality Be Considered Complete?" \cite{epr_paper}. Einstein, Podolsky, and Rosen challenged the foundations of quantum mechanics by pointing out a paradox. They argued that there exist pairs of particles, called EPR pairs, whose states are correlated in such a way that measuring a property of one particle automatically determines the property of the other particle, even if the particles are separated by a large distance. This strange behavior, they argued, could only be explained by "action at a distance," which violates the principle of locality. To counter this thought, Einstein came up with the idea of a hidden variable, which is an inaccessible property every particle holds. Later on, John Bell came up with a hypothetical experiment to determine the existence of hidden variables in particles \cite{bell-1964}. He demonstrated that his formulated inequality must hold true in the case when locally hidden variables exist. On the contrary, recent experiments have shown to violate the inequality \cite{aspect-1982}. \\
Several recent studies have demonstrated a positive outlook on the potential merger of chaos and quantum key distribution(QKD).
 \cite{8267797} proposes a technique to transmit data using quantum key distribution, OTP, and Huffman encoding. \cite{spitz2021chaos} investigates that the security of data transmission can be ensured using the synchronization chaos parameters. \cite{qkd_mahmud} research explores the utilization of chaotic communication through FSO technology, along with the sharing of chaos parameters via the BB84 quantum channel. By employing this approach, multiple layers of security are established, resulting in enhanced resilience and strengthened security. Moreover, the suggested method is well-suited for long-distance communication. Their work relies on QKD to transmit related Lorentz parameter values. However, our work uses Ekerts protocol \cite{ekert-1991} to generate parameters on both transmitter-receiver end. \cite{singh-2010} employs pulse position modulation to transmit chaotic signals.
 
\section{Preliminaries}\label{Sec:Preliminaries}

\subsection{Chaos Based Encryption}
Chaos based cryptography can be utilized in many different forms, but almost all of them involve the usage of a shared key between two parties. In this case, the key is a set of parameters. Using identical set of parameters, users on both end can generate identical key strings. Logistic map, henon map. lorentz map, and many other chaos maps are used from encryption all the time. Although most modern security relies on SHA-256 encryption algorithm, chaos maps are still widely used. Many image encryption schemes utilize the chaos to alter the pixel values \cite{spitz2021chaos} \cite{liu2012image} \cite{janani2021secure}.

\subsection{Logistic Map}
Logistic map is a one-dimensional chaos map that has been widely used in cryptography due to its computational efficiency. 
The logistic map equation is given by:

\begin{equation}
\label{eq:logistic map}
x_{n+1} = rx_n(1-x_n)
\end{equation}

In equation \ref{eq:logistic map}, $x_n$ represents the current value of the system at iteration $n$, and $x_{n+1}$ is the subsequent value obtained by applying the logistic map equation. The parameter $r$ controls the behavior of the system and plays a critical role in determining the nature of the resulting sequence.

\begin{figure}[!ht]
\includegraphics[width=.8\linewidth, height=18em]{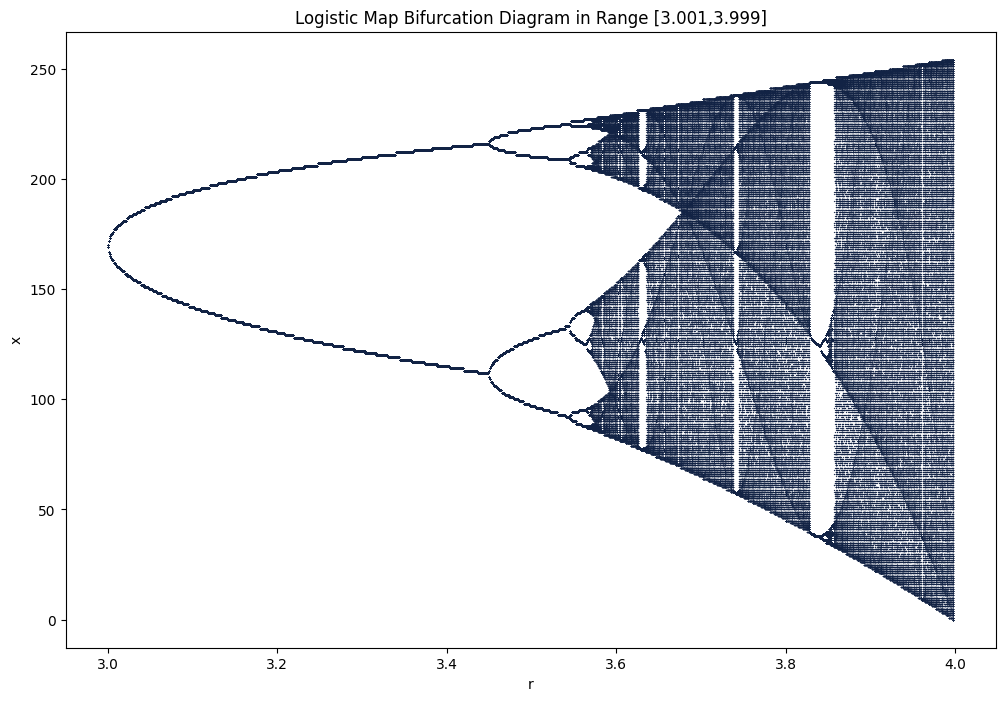}
\centering
\caption{Bifurcation diagram of Logistic Map.}
\label{bifurcation_diagram}
\end{figure} 
When $r$ takes on different values, the logistic map exhibits a range of behaviors. For specific values of $r$, the system settles into a stable state, converging to a single value. However, the logistic map displays chaotic dynamics for other values, particularly in the $3.57 < r < 4$. In this chaotic regime, the sequence generated by the logistic map is highly sensitive to initial conditions and exhibits complex, unpredictable behavior. Figure \ref{bifurcation_diagram} shows a bifurcation diagram of a logistic map. To maximize the chaos, one can use the $r$ values where chaos is nearly evenly distributed within the range [0,255].

\subsection{Ekert's E91 key distribution protocol}
Ekert's \cite {ekert-1991} relies on the principles of quantum entanglement, the no-cloning theorem, and the uncertainty principle to ensure the security of the key exchange. It introduces the use of Bell states to generate keys between two users, Alice and Bob. However, in order to facilitate this key generation process, we need the assistance of another user named Charlie. Charlie's role is to generate Bell states and distribute singlet particles to both Alice and Bob. The extended explanation of this process is as follows: \\
\begin{itemize}
    \item [ 1:] Bell State Generation: Charlie initiates the process by generating pairs of maximally entangled particles known as Bell states. These states are created in such a way that the measurement outcomes of the particles are correlated. Let's say we define
    \begin{equation}
       |B_{mn}\rangle_{AB}  =  \frac{1}{\sqrt{2}}(|0\rangle \otimes |m\rangle + (-1)^n|1\rangle \otimes |1 \oplus m\rangle)
    \end{equation}
where $m,n \in \{0,1\}$ and we can define four bell states as:

    \begin{equation}
    |\phi_+\rangle = |B_{00}\rangle ,  |\phi_-\rangle = |B_{01}\rangle,  
    |\psi_+\rangle = |B_{10}\rangle ,  |\psi_-\rangle = |B_{11}\rangle
    \end{equation}

    \item[ 2:] Singlet Distribution: Once the Bell states are created, Charlie distributes one particle from each entangled pair to both Alice and Bob. These particles are referred to as singlet particles.

    \item[ 3:] Measurement: Alice and Bob, who are geographically separated, receive their respective singlet particles. They independently perform measurements on their singlet particles using quantum measurement devices. The measurement outcomes are either 0 or 1 on the chosen basis, which is random and unpredictable due to the nature of quantum mechanics \cite{ekert-1991}. Measurement gates are determined using azimuthal angles and perpendicular to the trajectory of the particle.($\phi_1^a = 0$, $\phi_2^a = \frac{1}{4}\pi$,$\phi_3^a = \frac{1}{2}\pi$,$\phi_1^b = \frac{1}{4}\pi$,$\phi_2^b = \frac{1}{2}\pi$, $\phi_3^b = \frac{3}{4}\pi$)
    \item[ 4:] Correlated Measurement Results: After performing the measurements, Alice and Bob communicate their measurement results to each other using a classical channel. They compare their basis choices. If their basis choices match, they use the corresponding measurement outcomes to form a key. If the bases don't match, the data from that trial is discarded. They compare their measurement outcomes for each pair of singlet particles.
    
    \item[ 5:] Key Extraction: Alice and Bob apply a predetermined rule to correlate their measurement results. Based on this correlation, they extract a shared key. The shared key is a string of bits that can be used for secure communication.
    
\item [ 6:] Security Check: The security of the E91 QKD protocol relies on the ability to detect the presence of an eavesdropper. If an eavesdropper is detected, the key generation process can be aborted, ensuring the security of the key. By involving Charlie and utilizing the concept of entangled particles, Alice and Bob can generate a shared key pair without directly exchanging information. The use of Bell states and singlet particles adds an extra layer of security to the key generation process, making it resistant to eavesdropping attempts. 
\end{itemize} 
\subsection{Eavesdropper Detection using CHSH inequality}
In order for Alice and Bob to detect the presence of an anomaly, such as an eavesdropper, they both can take two steps. In preparing and measuring key distribution schemes, Alice and Bob share part of their key to look for any potential mismatch. If found, they assume there was an anomaly. Entanglement-based protocols such as E91 offer an additional test, where Alice and Bob calculate whether they violate the CHSH inequality\cite{chsh} or not. If not violated, they can assume there was an anomaly.
Ekert introduced an extra basis to enable the direct detection of an eavesdropper's presence without compromising the confidentiality of the key. 
The correlation coefficient between Alice's (\(a_i\)) and Bob's (\(b_j\)) measurements is expressed as:
 \begin{equation}
 E(a_i,b_j) = P_{++}(a_i,b_j) + P_{--}(a_i,b_j) - P_{+-}(a_i,b_j) - P_{-+}(a_i,b_j) 
 \end{equation}
A quantity denoted as \(S\) is defined as the sum of all correlation coefficients for which Alice and Bob employed detectors with differing orientations (incompatible bases):
\begin{equation}    
S = E(a_1, b_1) - E(a_1,b_3) + E(a_3,b_1) + E(a_3,b_3)
\end{equation} 
In order for local realism to be valid, Bell\cite{bell-1964} proved that the value of $-2\leq S
\leq 2$. However, quantum mechanics gives that $S = 2\sqrt{2}$. In our work, we have initialized our bell pairs using \begin{math}|\phi_+\rangle\end{math} state.

CHSH inequality is widely used as mean to construct a mathematical inequality used to test local realism. Local realism consists of ideas like locality and universal realism, claiming that properties of objects in the universe exist irrespective to our observation. Einstein \cite{epr_paper} argued the existence of entanglement claiming that it violated locality. However, CHSH inequality was introduced following Bell's original theorem \cite{chsh}, showing a path to test this theorem in reality. \cite{aspect-1982} tested whether this inequality holds true in quantum realm and found that it doesn't, which resulted in a Nobel prize in 2022. \\ 
In classical realm, the inequality shall be:
\begin{equation} \label{Eq:inequality}
-2\leq S\leq 2
\end{equation}
We can obtain our inequality following the previous correlation observations: 
\begin{equation}\label{Eq:ineqality_violation}
    \langle S\rangle = \langle x\otimes w\rangle_{\phi^+} - \langle x\otimes v\rangle_ {\phi^+} + \langle z\otimes v\rangle_ {\phi^+} + \langle z\otimes w\rangle_ {\phi^+}  = 2\sqrt{2}
\end{equation}
Equation \ref{Eq:ineqality_violation} clearly violates equation \ref{Eq:inequality}. In our experiment, we can expect to get a chsh value close to $2\sqrt{2}$, otherwise we can claim there were some form of anomaly like an eavesdropper.

\section{Proposed Architecture}\label{Sec:Architecture}
\begin{figure*}[!ht]
    \centering
    \includegraphics[width=\linewidth, height = 27em]{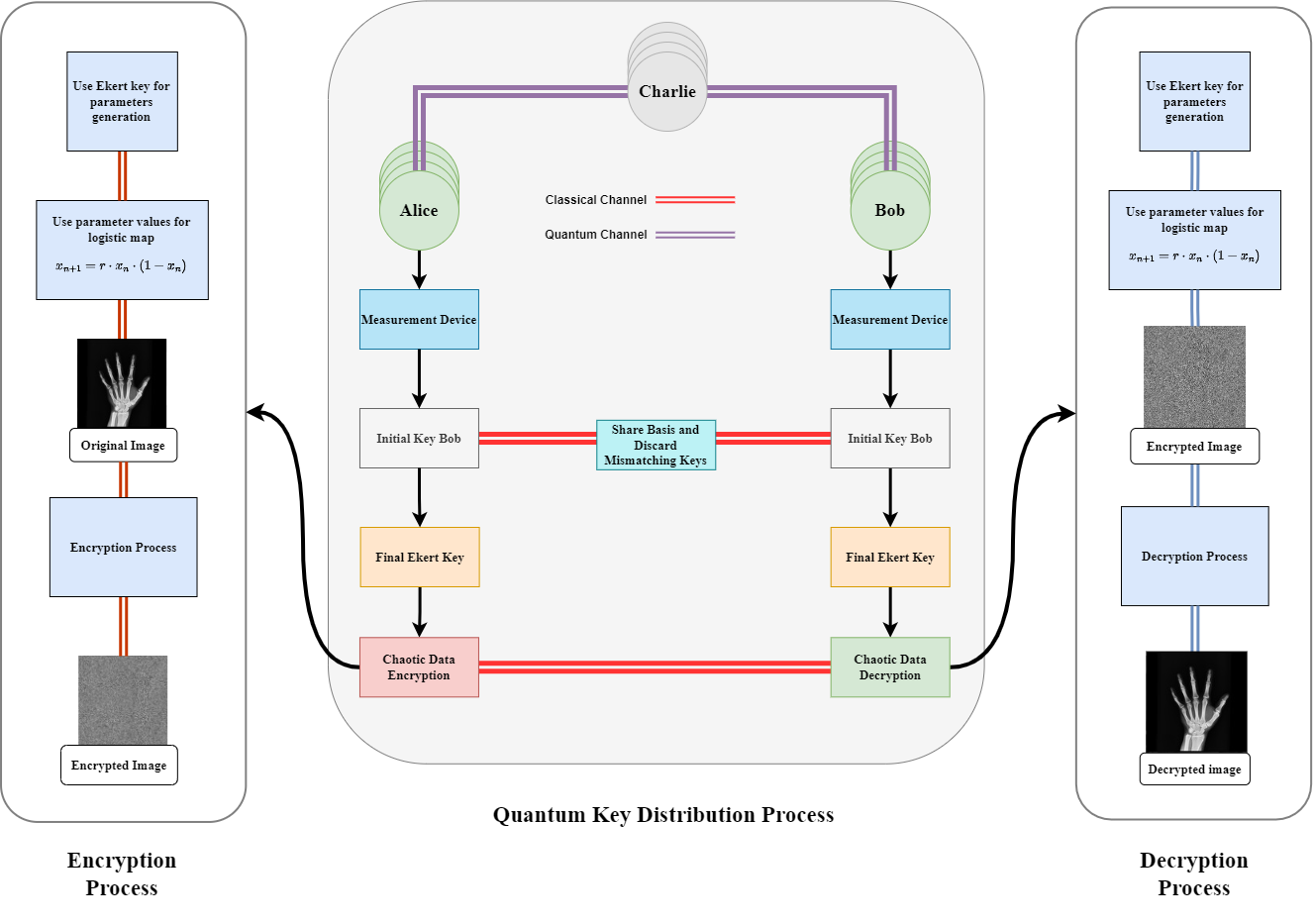}
    \centering
    \caption{
elucidates the proposed study framework, which involves the distribution of entangled particles from Charlie(Singlet source) to Alice and Bob. Upon reception, Alice and Bob employ measurement devices to generate random measurement bases and initial keys, which are compared to derive the Final key. This final key is subsequently utilized in chaotic data encryption and decryption phases, where the Ekert key(final key) generates Logistic map parameters for facilitating image encryption and decryption}
    \label{Fig:methodology}
\end{figure*}

Our architectural approach transforms security in an adaptive fashion through the utilization of the pure randomness achieved by the E91 Protocol. This strategy ensures the distribution of keys across the encryption and decryption endpoints. In Figure \ref{Fig:methodology}, we provide a concise overview of our proposed design, commencing with the initial block detailing the Quantum Key Distribution (QKD) process.

\subsection{Initialization}
\begin{figure}[!ht]
    \centering
    \includegraphics[width=\linewidth, height=8em]{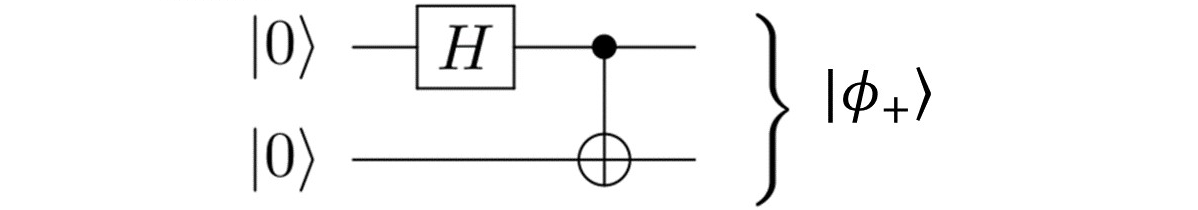}
    \caption{Singlet circuit for creating entanglement qbits}
    \label{fig:SigletCircuit}
\end{figure}

Following the procedure outlined by Ekert (1991) \cite{ekert-1991}, referred to as Charlie, a user shares a set of entangled qubits, denoted as $n$, with both Alice and Bob by using the circuit shown in Figure \ref{fig:SigletCircuit}. 

\subsection{Measurement}
\begin{figure}[!ht]
\includegraphics[width=\linewidth, height = 16em]{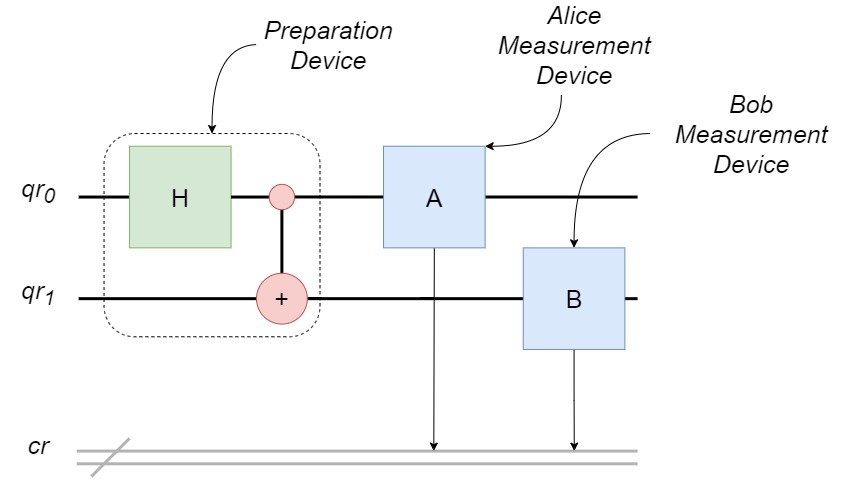}
\centering
\caption{Combined circuits for E91 protocol's Key distribution.}
\label{Combined_Circuit}
\end{figure}

\begin{figure}[!ht]
    \centering
    \includegraphics[width=.7\linewidth, height=14em]{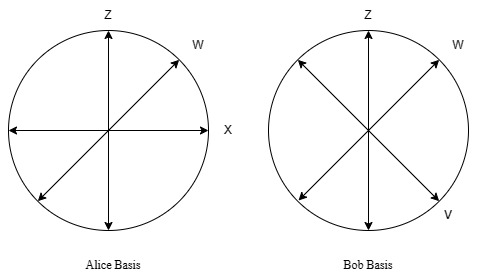}
    \caption{The polarization of $X$, $W$, $Z$, and $V$ observables}
    \label{fig:polarized}
\end{figure}

\begin{figure*}[h!] 
\begin{subfigure}{\textwidth}
    \centering
    \includegraphics[width=0.98\textwidth]{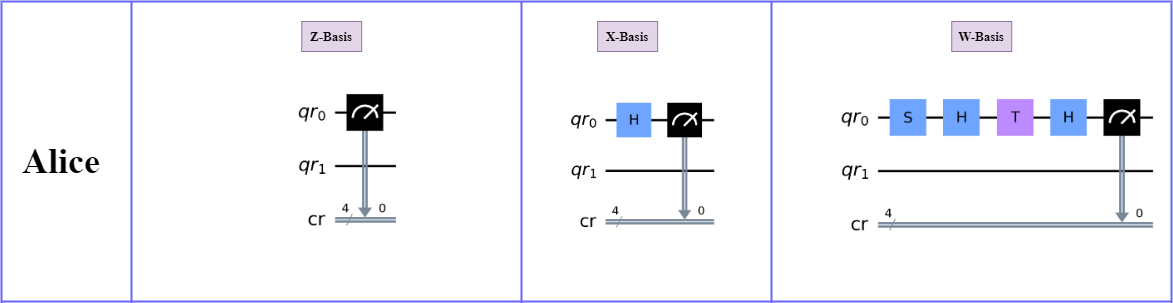}
    \caption{Alice's measurement circuits on $Z$, $X$, $W$ observables}
    \label{fig:Alice_measurements}
\end{subfigure}
\begin{subfigure}{\textwidth}
    \centering
    \includegraphics[width=0.98\textwidth]{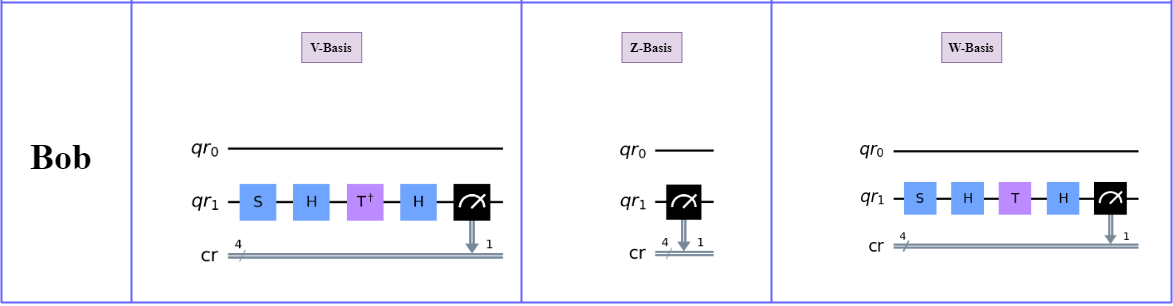}
    \caption{Bob's measurement circuits on $V$, $Z$, $W$ observables}
    \label{fig:Bob_measurements}
\end{subfigure}
        
\caption{Alice and Bob measurement circuits.}
\label{fig:figures}
\end{figure*}

Alice and Bob independently select arrays of $n$ random measurement basis denoted as $B_a$(Shown in Figure \ref{fig:Alice_measurements}) and $B_b$(Shown in Figure \ref{fig:Bob_measurements}), respectively, based on $X$, $W$, $Z$, and $V$(Shown Figure \ref{fig:polarized})in observables. 
Subsequently, they perform measurements based on $B_a$ and $B_b$ on the received qubits at both ends(Shown in Figure \ref{Combined_Circuit}), yielding an initial key. After their measurement, Alice and Bob proceed to share their respective basis arrays($B_a$ and $B_b$) through a classical communication channel. This step involves a thorough comparison of these arrays, enabling them to identify and discard keys that lack a common basis. It is worth emphasizing that the information shared through the classical channel pertains to their basis arrays, not the original encryption keys. This process ultimately results in Alice and Bob both possessing identical encryption keys.

\subsection{Parameter generation for Chaotic System}
\begin{figure}[!ht]
\includegraphics[width=.6\linewidth, height = 18em]{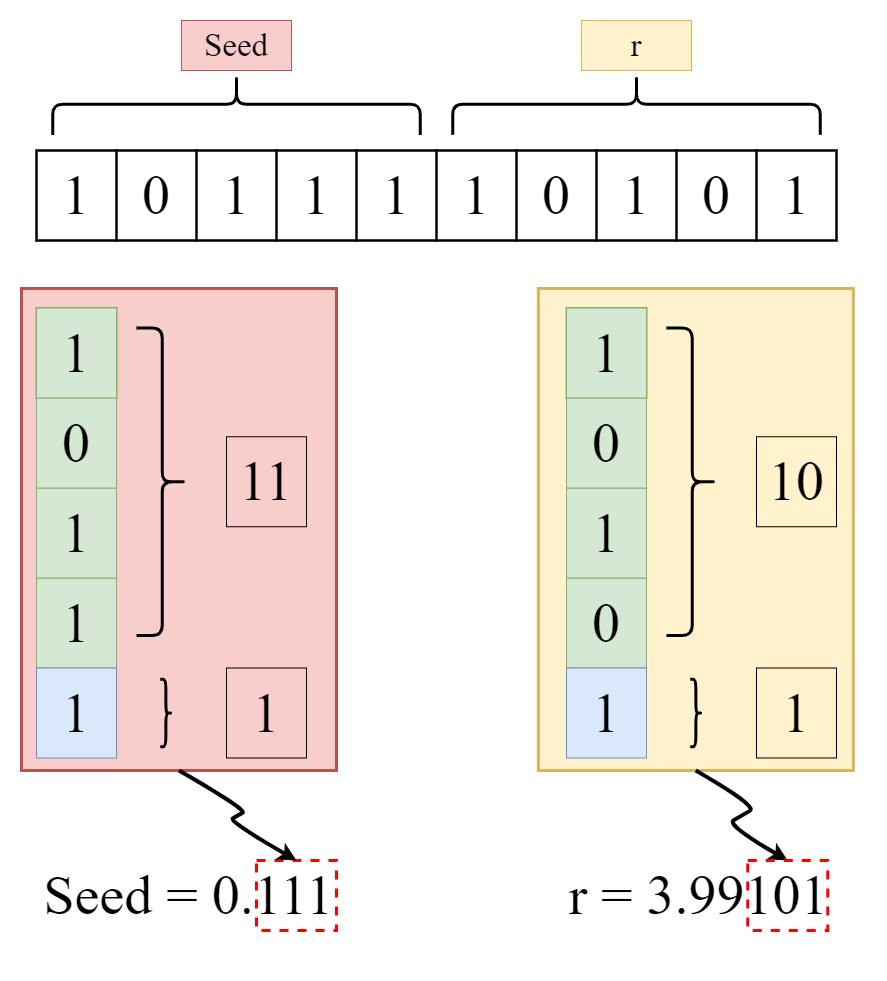}
\centering
\caption{ elucidates an example visual representation of the key splitting process. Initially, the first half of the key is divided, with each quartet of digits transformed into their respective decimal values. Subsequently, these values are recombined to produce the value of the $x_0$ parameter. In a similar process, we extract the value of $r$. Assuming that both Alice and Bob possess identical, unaltered keys, they can consistently generate the same initial conditions for a functional logistic map.}
\label{Fig:key_division}
\end{figure}

By utilizing the shared keys, Alice takes several crucial steps to kick-start an encryption scheme, in our case, the logistic chaos encryption approach. To do so, they employ the shared key to derive essential parameters such as the initial condition ($x_0$) and reproduction value ($r$). The Algorithm \ref{Algo:logistic_chaos_Gen} depicts the straightforward process of generating $X_0$ and $r$ values. These parameters are achieved by dividing the key into two equal halves, with one half being designated for the $r$ and the other half for the $x_0$. To ensure the $x_0$ falls within the range of [$0, 1$], and considering that the system exhibits heightened chaotic behavior when the parameter $r$ lies between [$3.54$,$4$], particularly when it approaches $3.99$, we have come up with an approach to incorporate the key immediately after the decimal points shown in Figure \ref{Fig:key_division}. Notably, for the parameter $r$, we have observed an intensified state of chaos in proximity to the value $3.99$, which led us to add the key right after the $3.99$ value. 
The logistic map offers the capability of yielding varying degrees of chaos by making minor adjustments to a single digit within a specific range.
\begin{algorithm}[!ht]
\caption{Generate Logistic Chaos Parameters}\label{Algo:logistic_chaos_Gen}
\begin{algorithmic}[1]
\Function{GenerateLogisticChaosParam}{$key$}
    \State $decimal\_list \gets keyToDecimal(key)$
    \State $seed \gets generateSeed(decimal\_list)$
    \State $r \gets generateRValue(decimal\_list)$
    
    \If{$seed = 0$}
        \State $seed \gets 0.33$
    \EndIf
    
    \If{$r > 4$}
        \State $r \gets 3.999999$
    \EndIf
    
    \State \textbf{return} $decimal\_list, seed, r$
\EndFunction
\end{algorithmic}
\end{algorithm}

Utilizing these initial conditions, Alice and Bob both generate an array of key streams shown in algorithm \ref{algo:generate_key_stream}. 

\begin{algorithm}[!ht]
\caption{Generate Key Stream}
\label{algo:generate_key_stream}
\begin{algorithmic}[1]
\Function{generate\_key\_stream}{$\text{seed}, \text{length}, \text{r}$}
  \State $key\_stream \gets$ []
  \State $x \gets$ seed 
  \For{$i\gets 1$ to length}
    \State $x \gets$ \Call{logistic\_map}{$x, r$}
    \State Append ($\text{int}(x) \times 256$) to $key\_stream$
  \EndFor
  \State \Return $key\_stream$
\EndFunction
\end{algorithmic}
\end{algorithm}

\subsection{Data Encryption \& Decryption}
 Our encryption follows pixel-wise image encryption method. Initially, Alice converts her image to a grey scale image, as it is suitable for working with a  one-dimensional(1D) map such as a logistic one.
Following the conversion, Alice uses her key stream elements to perform $XOR$ with each pixel value ranging from [$0,255$]. This allows the encrypted pixels to be distorted from their original states. The Algorithm \ref{Algo:encrypt_decrypt} shows the encryption process. After successful encryption, Alice will send her encrypted image through some classical channel. Other eavesdropper parties cannot decrypt this since they do not hold the parameter values. The only possible way to decrypt is via brute force attack, where they will have to try around 10 trillion combinations of parameter values. After the conversion process, Alice obtains a key stream to conduct an $XOR$ operation with every pixel value, encompassing the range of [0, 255]. This operation introduces a distortion into the encrypted pixels, rendering them unrecognizable compared to their original states. Once the encryption is successfully executed, Alice transmits the encrypted image through a classical communication channel. Unauthorized eavesdropping parties are detected in their attempts to decipher the image, as they lack the knowledge of parameter values. The only conceivable method available to them for decryption entails a brute force approach, entailing the testing of approximately 10 trillion parameter value combinations. By implementing this system design, we achieve secure encryption and decryption without the need to share keys, ensuring the confidentiality of the data transmission. 

\begin{algorithm}[!ht]
\caption{Encrypt or Decrypt Image}\label{Algo:encrypt_decrypt}
\begin{algorithmic}[1]
\Function{EncryptDecryptImage}{$image, key\_stream$}
    \State $encrypted\_image \gets \text{Copy}(image)$
    \State $height, width \gets \text{Shape}(encrypted\_image)$
    \State $i \gets 0$
    
    \For{$x$ \textbf{in} $\text{Range}(height)$}
        \For{$y$ \textbf{in} $\text{Range}(width)$}
            \State $encrypted\_image[x, y] \gets encrypted\_image[x, y] \oplus key\_stream[i \mod \text{Length}(key\_stream)]$
            \State $i \gets i + 1$
        \EndFor
    \EndFor
    
    \State \textbf{return} $encrypted\_image$
\EndFunction
\end{algorithmic}
\end{algorithm}

\subsection{Eavesdropping Protection}
Since our entire scheme is built upon chaos maps, they are always initial condition dependent. Any minor change to the initial condition value will alter the value of the E91 key, resulting in distortion while decryption. For example, we have flipped first key of E91 key (which is an input of initial key generation function) and the resulted decryption was unsuccessful. However, another way to detect eavesdropper it to compute the CHSH correlation value discussed in the preliminaries.\\

\begin{figure}[!ht] 
\includegraphics[width=.9\linewidth, height = 15em]{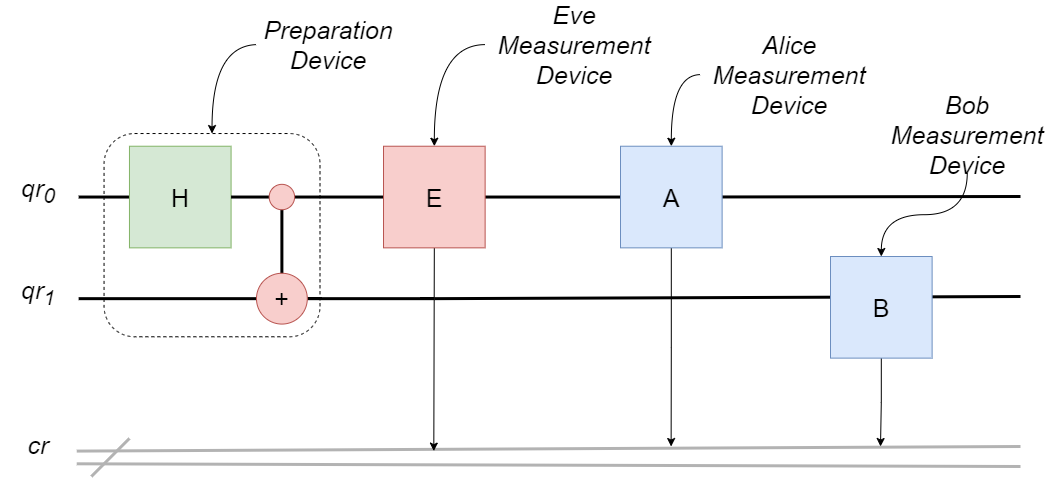}
\centering
\caption{Eavesdropper Eve interfering with Alice's end of the communication}
\label{Fig:eavesdropper}
\end{figure}

Figure \ref{Fig:eavesdropper} illustrates a potential eavesdropper, Eve, interfering with Alice's end of the communication. Eve intercepts the quantum channel, measures the bits sent to Alice, and sends her a randomly prepared state. This action collapses the state of the particle for Alice, breaking the maximal entanglement between the two particles. Consequently, the CHSH correlation value of \(|S| = 2\sqrt{2}\) is no longer achievable.

\section{Implementation Details}\label{Sec:Implementation}
Our research employed a combination of powerful tools and libraries to implement our proposed scheme effectively. IBM's Qiskit \cite{Qiskit} served as the backbone for developing and simulating quantum circuits, enabling the implementation of the E91 protocol. NumPy's \cite{NumPyoliphant2006guide} extensive numerical computation capabilities facilitated the handling of large arrays and complex calculations associated with the logistic map. Additionally, OpenCV's \cite{CV2itseez2014theopencv} image processing functionalities proved invaluable for grayscale conversion. 
\section{Result Analysis}\label{Sec:results}
\subsection{Key-space and Key Sensitivity Analysis}
The key space for a logistic chaotic map is theoretically infinite. This is because both $seed(x_0)$ and $r$ can be any real number between 0 and 1. However, in practice, the key space is limited by the computer's precision used to implement the algorithm. For example, if a computer uses $64-bit$ floating-point numbers\cite{Floating_Point}, then the precision of $x0$ and $r$ will be limited to $2^{64}$ possible values. This would mean that the key space would be approximately $1.8 \times 10^{19}$.

\begin{figure*}[!ht]
\centering
\begin{subfigure}{0.3\textwidth}
    \includegraphics[width=\textwidth]{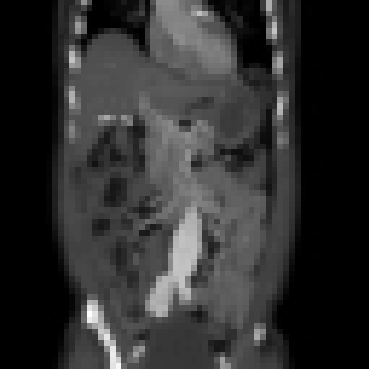}
    \caption{Original Image($64\times64$)}
    \label{fig:abdomine}
\end{subfigure}
\hfill
\begin{subfigure}{0.3\textwidth}
    \includegraphics[width=\textwidth]{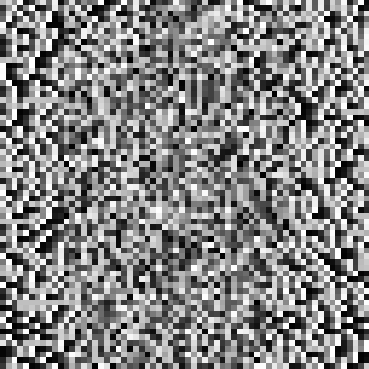}
    \caption{Encrypted Image($64\times64$)}
    \label{fig:abdomine_encrypted}
\end{subfigure}
\hfill
\begin{subfigure}{0.3\textwidth}
    \includegraphics[width=\textwidth]{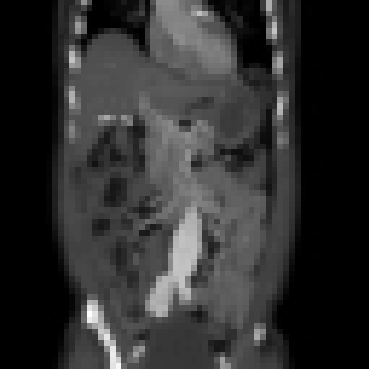}
    \caption{Decrypted Image($64\times64$)}
    \label{fig:abdomine_decrypted}
\end{subfigure}
\hfill
\begin{subfigure}{0.3\textwidth}
    \includegraphics[width=\textwidth]{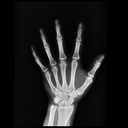}
    \caption{Original Image($128\times128$)}
    \label{fig:xray}
\end{subfigure}
\hfill
\begin{subfigure}{0.3\textwidth}
    \includegraphics[width=\textwidth]{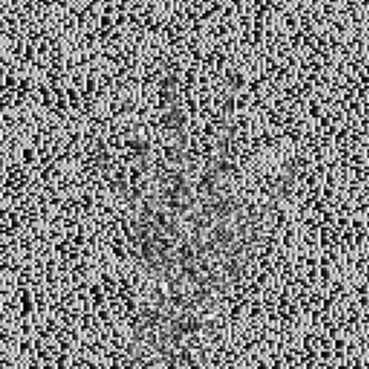}
    \caption{Encrypted Image($128\times128$).}
    \label{fig:xray_encrypted}
\end{subfigure}
\hfill
\begin{subfigure}{0.3\textwidth}
    \includegraphics[width=\textwidth]{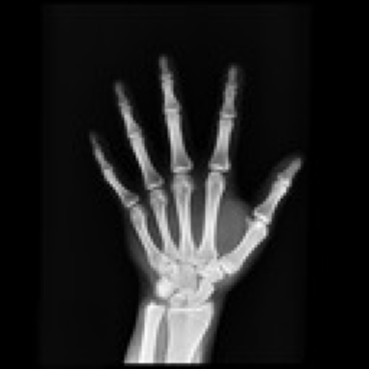}
    \caption{Decrypted Image($128\times128$)}
    \label{fig:xray_decrypted}
\end{subfigure}
\hfill
\begin{subfigure}{0.3\textwidth}
    \includegraphics[width=\textwidth]{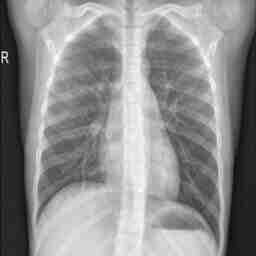}
    \caption{Original Image($256\times256$)}
    \label{fig:first}
\end{subfigure}
\hfill
\begin{subfigure}{0.3\textwidth}
    \includegraphics[width=\textwidth]{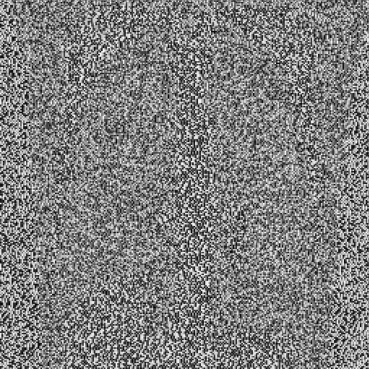}
    \caption{Encrypted Image($256\times256$)}
    \label{fig:chest_encrypted}
\end{subfigure}
\hfill
\begin{subfigure}{0.3\textwidth}
    \includegraphics[width=\textwidth]{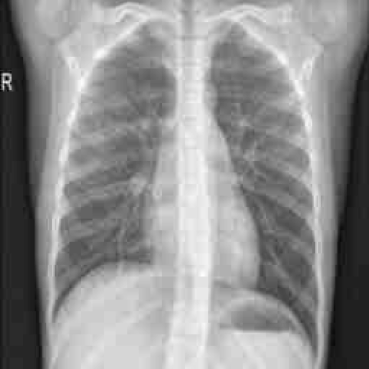}
    \caption{Decrypted Image($256\times256$)}
    \label{fig:chest_decrypted}
\end{subfigure}
        
\caption{Depicts successful encryption across various image pixels}
\label{fig:Success_Encryption}
\end{figure*}

\begin{figure*}[!ht]
\centering
\begin{subfigure}{0.3\textwidth}
    \includegraphics[width=\textwidth]{abdomine.png}
    \caption{Original Image($64\times64$)}
    \label{fig:Abdomine2}
\end{subfigure}
\hfill
\begin{subfigure}{0.3\textwidth}
    \includegraphics[width=\textwidth]{abdomine_encrypted.png}
    \caption{Encrypted Image($64\times64$)\\ seed: 0.1110540121\\ r: 3.99620155131585}
    \label{fig:abdomine_encrypted_1}
\end{subfigure}
\hfill
\begin{subfigure}{0.3\textwidth}
    \includegraphics[width=\textwidth]{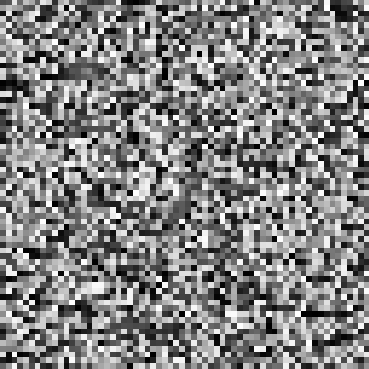}
    \caption{Failed decryption($64\times64$)\\ seed: 0.3105401212\\ r :3.99620155131585}
    \label{fig:abdomine_failed_decrypted}
\end{subfigure}
\hfill
\begin{subfigure}{0.3\textwidth}
    \includegraphics[width=\textwidth]{xray.jpg}
    \caption{Original Image($128\times128$)}
    \label{fig:Xray_1}
\end{subfigure}
\hfill
\begin{subfigure}{0.3\textwidth}
    \includegraphics[width=\textwidth]{xray_encrypted.png}
    \caption{Encrypted Image($128\times128$)\\ seed: 0.1110540121\\ r: 3.99620155131585}
    \label{fig:xray_encrypted_1}
\end{subfigure}
\hfill
\begin{subfigure}{0.3\textwidth}
    \includegraphics[width=\textwidth]{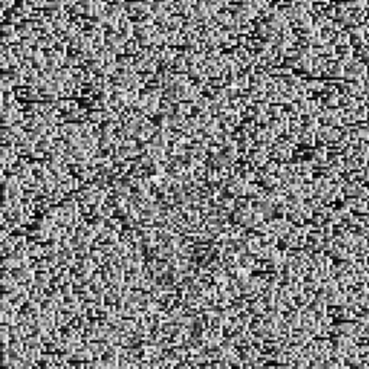}
    \caption{Failed decryption($128\times128$)\\ seed: 0.3105401212\\ r: 3.99620155131585}
    \label{fig:xray_failed_decrypted}
\end{subfigure}
\hfill
\begin{subfigure}{0.3\textwidth}
    \includegraphics[width=\textwidth]{chest.jpg}
    \caption{Original Image($256\times256$)}
    \label{fig:chest_1}
\end{subfigure}
\hfill
\begin{subfigure}{0.3\textwidth}
    \includegraphics[width=\textwidth]{chest_encrypted.png}
    \caption{Encrypted Image($128\times128$)\\ seed: 0.1110540121\\ r: 3.99620155131585}
    \label{fig:chest_encrypted_1}
\end{subfigure}
\hfill
\begin{subfigure}{0.3\textwidth}
    \includegraphics[width=\textwidth]{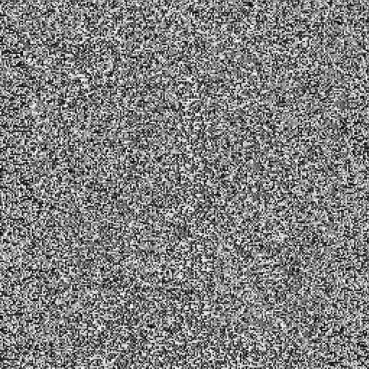}
    \caption{Failed decryption($256\times256$)\\ seed: 0.3105401212\\ r: 3.99620155131585}
    \label{fig:chest_failed_decrypted}
\end{subfigure}
\caption{Depicts failed decryption across various image pixels}
\label{Fig:Failed_Decryption}
\end{figure*}

In our works, we have used 500 entangled qubits. Therefore, Alice and Bob both receive a key string length of 500 bits. For a critical length of 500, we could use $250$ bits to represent $x_0$ and $250$ bits to represent $r$. This would mean that the key space would be approximately $2^{500}$, which is approximately $1.1 \times 10^{150}$. This is a very large number, and it is unlikely that any computer can ever store or process a key of this size. However, knowing the theoretical limits of the key space for this scheme is still helpful.
Our research has demonstrated Alice's successful application of the encryption process and subsequent decryption by Bob, leading to the secure and reliable transmission of the image. This achievement serves as a testament to the effectiveness and validity of the proposed cryptographic scheme. The experimental process involved the use of a random $seed$ and a $r$-value generated from quantum E91 distribution, which was crucial in establishing the encryption and decryption parameters. Alice encrypts the image data using the specified parameter values. On the receiving end, Bob successfully reverses the encryption process using the same $seed$ and $r$-value as Alice's. Figure \ref{fig:Success_Encryption} shows an image encryption using $seed$ and $r$-value generated from an arbitrary E91 distribution. Alice encrypts the image using the mentioned parameter values, and Bob follows by decrypting the image using the same parameter values. On the other hand, Figure \ref{Fig:Failed_Decryption} refers to the case where Bob did not have the same E91 key stream as Alice. In this case, Bob used his keys to generate chaos parameters in order to decrypt the image, but he failed as the chaos parameters are slightly different than that of Alice. Since chaotic maps are very sensitive to initial values, the resulting integers widely differed, resulting in a failed decryption process.

\subsection{Evesdropping and CHSH value Evaluation}
we have measured the CHSH-inequality values for maximally entangled qubits used in E91 procedure in Table.\ref{table:CHSH test}. Alongside Table.\ref{table:CHSH test eavesdropper} demonstrates the violation of CHSH inequality when the qubits are not maximally entangled in the case of an eavesdropper.  
 
\begin{table}[!ht]
\centering
\caption{CHSH correlation value test}
\label{table:CHSH test}
\begin{tabularx}{0.8\textwidth} { 
   >{\raggedright\arraybackslash}X
   >{\centering\arraybackslash}X
   >{\raggedleft\arraybackslash}X}
 \toprule
 \textbf{Iterations}  & \textbf{Number of bits} & \textbf{CHSH Correlation}   \\
 \midrule
  1  & 106  & 2.797  \\
\midrule
2  & 104  &  2.929  \\
\midrule
 3  & 107  & 2.911  \\
\bottomrule
\end{tabularx}
\end{table}

\begin{table}[!ht]
\centering
\caption{CHSH correlation value in Eavesdropping scenario}
\label{table:CHSH test eavesdropper}
\begin{tabularx}{0.8\textwidth} { 
   >{\raggedright\arraybackslash}X
   >{\centering\arraybackslash}X
   >{\raggedleft\arraybackslash}X
   >{\raggedleft\arraybackslash}X
      >{\raggedleft\arraybackslash}X}
 \toprule
 \textbf{Iterations} & \textbf{Number of Mismatching bits} & \textbf{Number of Mismatching bits} & \textbf{CHSH correlation} & \textbf{Eves knowledge of Alice's key} \\
 \midrule
1  & 101 & 13 & 0.973 & 87\%  \\
\midrule
2  & 112 & 18  & 0.579 & 84\%  \\
\midrule
3  & 1 & 22  & 1.467 & 82\%  \\
\bottomrule
\end{tabularx}
\end{table}

\subsection{Histogram Analysis}
\begin{figure*}[!ht]
\centering
\begin{subfigure}{0.4\textwidth}
    \includegraphics[width=\textwidth]{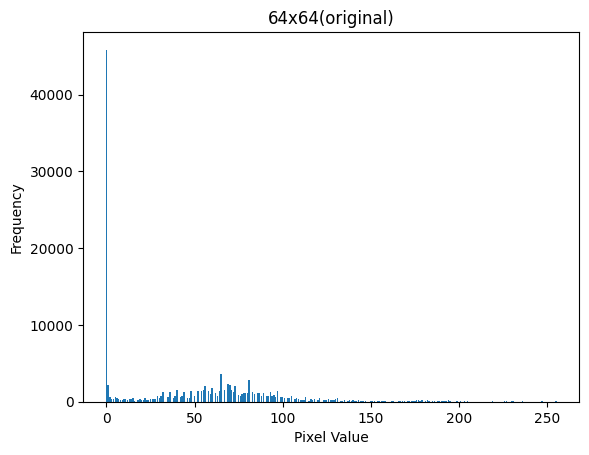}
    \caption{Original Image($64\times64$)}
    \label{fig:64o}
\end{subfigure}
\hfill
\begin{subfigure}{0.4\textwidth}
    \includegraphics[width=\textwidth]{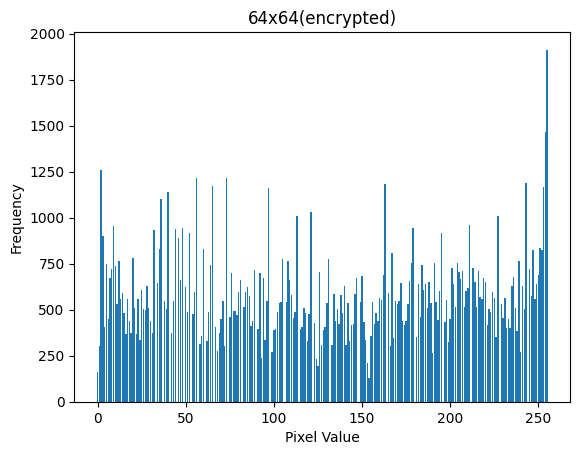}
    \caption{Encrypted Image($64\times64$)}
    \label{fig:64e}
\end{subfigure}
\hfill
\begin{subfigure}{0.4\textwidth}
    \includegraphics[width=\textwidth]{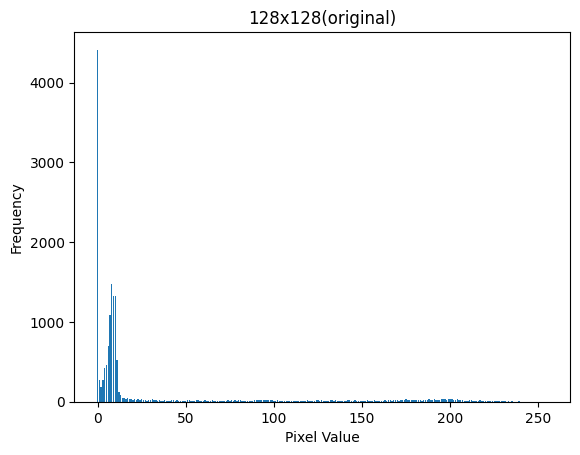}
    \caption{Original Image($128 \times128$).}
    \label{fig:third}
\end{subfigure}
\hfill
\begin{subfigure}{0.4\textwidth}
    \includegraphics[width=\textwidth]{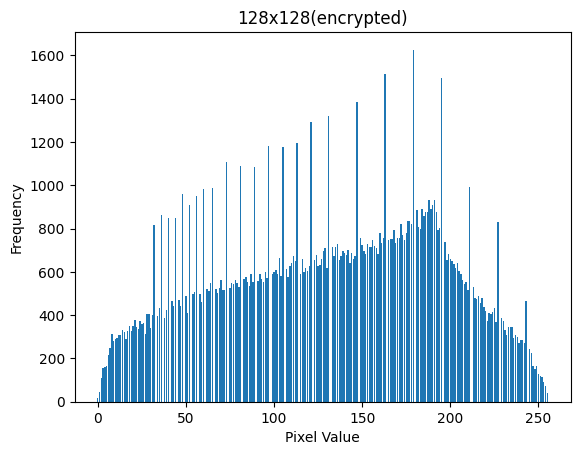}
    \caption{Encrypted Image($128 \times128$).}
    \label{fig:128e}
\end{subfigure}
\hfill
\begin{subfigure}{0.4\textwidth}
    \includegraphics[width=\textwidth]{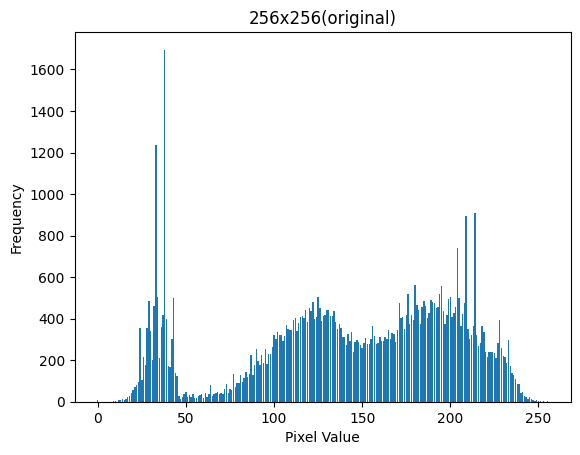}
    \caption{Original Image($256\times256$)}
    \label{fig:256o}
\end{subfigure}
\hfill
\begin{subfigure}{0.4\textwidth}
    \includegraphics[width=\textwidth]{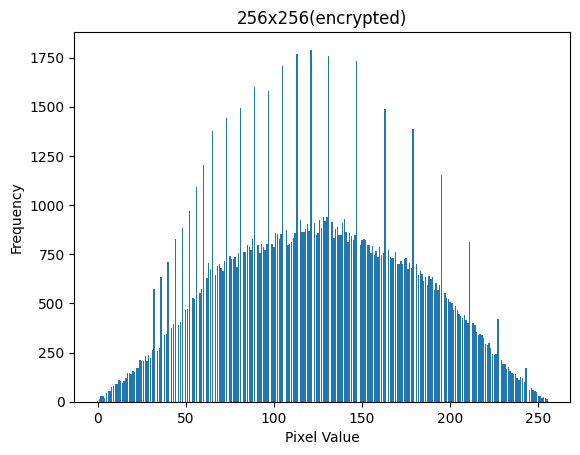}

    \caption{Encrypted Image($256 \times256$).}
    \label{fig:256e}
\end{subfigure}       
\caption{Histogram analysis of the images.}
\label{fig:Histogram}
\end{figure*}

The image histogram serves as a visual representation of the distribution of grayscale values within an image. To withstand statistical attacks, the encrypted image should exhibit a uniform grayscale distribution. This uniformity can be assessed by plotting the histograms of the cipher image. Figure \ref{fig:Histogram} illustrates the histograms of the input image and the corresponding encrypted image. It is evident from the figure that the histograms of the proposed encrypted images closely resemble a uniform distribution and exhibit significant alterations compared to the input image histogram.

\subsection{Information Entropy Analysis}
The Shannon entropy\cite{shannon1948} measures the uncertainty or randomness of a random variable. In information theory, it's defined as:
\begin{equation}
H(X) = - \sum_{i=1}^{n} P(x_i) \cdot \log_b(P(x_i))
\end{equation}
Where \( H(X) \) represents the entropy of the random variable \(X\). \( P(x_i) \) is the probability of the outcome \(x_i\) occurring. \( n \) is the number of possible outcomes.
In order to demonstrate the randomness generated inside the image itself, we have utilized the idea of entropy analysis. The average value of the proposed encryption method is closer to the ideal value of 8, as shown in Table \ref{table:entropy}. Consequently, the information entropy of the encrypted image approaches 8, indicating that a more secure cryptosystem leaves less information available for potential attackers.
\begin{table}[!ht]
\centering
\caption{Information entropy of the plain and encrypted images}
\label{table:entropy}
\begin{tabularx}{0.8\textwidth} { 
   >{\raggedright\arraybackslash}X
   >{\centering\arraybackslash}X
   >{\raggedleft\arraybackslash}X}
 \toprule
 \textbf{Images} & \textbf{Plain Image(s)} & \textbf{Encrypted Image(s)} \\
 \midrule
  $64\times64$  & 5.35  & 7.45  \\
\midrule
$128\times128$  & 4.88  & 7.39  \\
\midrule
 $256\times256$  & 7.51  & 7.22  \\
\midrule
\textbf{Average}  & \textbf{5.91}  & \textbf{7.35}  \\
\bottomrule
\end{tabularx}
\end{table}

\subsection{Execution Time Complexity}
The Time Complexity of the proposed encryption scheme is assessed by evaluating the time required for image encryption and regional data encryption. Table \ref{table:Time_complexity} summarizes our proposed scheme's encryption and decryption times. 
\begin{table}[!ht]
\centering
\caption{Execution Time complexity of the proposed scheme.}
\label{table:Time_complexity}
\begin{tabularx}{0.8\textwidth} { 
   >{\raggedright\arraybackslash}X
   >{\centering\arraybackslash}X
   >{\raggedleft\arraybackslash}X}
 \toprule
 \textbf{Images} & \textbf{Encryption Time(ns)} & \textbf{Decryption Time(ns)} \\
 \midrule
  $64\times64$  & 6.98  & 6.982  \\
\midrule
$128\times128$  & 40.18  & 29.6  \\
\midrule
 $256\times256$  & 177.28  & 162.66  \\
 \midrule
\textbf{Average}  & \textbf{74.81}  & \textbf{199.242}  \\
\bottomrule
\end{tabularx}
\end{table}

\subsection{Differential Attack Analysis}
The Normalized Pixel Change Rate(NCPR) \cite{NPCR} test evaluates the strength of a chaotic encryption algorithm that measures the percentage of pixels in the encrypted image that have changed compared to the original image. The higher the NPCR value, the stronger the encryption algorithm. Similarly, The Unified Average Changing Intensity(UACI) \cite{NCPR_UCAI} test is another method for evaluating the strength of a chaotic encryption algorithm. It measures the average intensity change between the original and encrypted images. The lower the UACI value, the stronger the encryption algorithm.
NPCR and UACI formulas can be given by,
\begin{equation}
NPCR = \frac{\sum_{i=1}^{H} \sum_{j=1}^{W} D(O_{ij}, C_{ij})}{H \times W} \times 100\%,
\end{equation}

\begin{gather*} 
\begin{cases}
  D(O_{ij}, C_{ij}) = 0;& \text{$for$ }O_{ij} = C_{ij}\\    
  D(O_{ij}, C_{ij}) = 1;& else
\end{cases}
\end{gather*}

\begin{equation}
UACI = \frac{\sum_{i=1}^{H} \sum_{j=1}^{W} \left| O_{ij} - C_{ij} \right|}{255 \times H \times W} \times 100\%,
\end{equation}

where $O_{ij}$ is the original image pixel value at position $(i, j)$ and $C_{ij}$ is the encrypted image pixel value at position $(i, j)$. $D(O_{ij}, C_{ij})$ is defined by the stated condition. $H$ is the height of the image. $W$ is the width of the image.\\
Table \ref{table:Diffatack} summarizes our proposed scheme's NCPR and UACI test results. 

\begin{table}[H]
\centering
\caption{NPCR and UACI of the encrypted images}
\label{table:Diffatack}
\begin{tabularx}{0.8\textwidth} { 
   >{\raggedright\arraybackslash}X
   >{\centering\arraybackslash}X
   >{\raggedleft\arraybackslash}X}
 \toprule
 \textbf{Images} & \textbf{NPCR(\%)} & \textbf{UACI(\%)} \\
 \midrule
  $64\times64$  & 99.66 & 35.32  \\
\midrule
$128\times128$  & 99.91  & 44.83 \\
\midrule
 $256\times256$  & 99.42  & 23.71  \\
\midrule
\textbf{Average}  & \textbf{99.66}  & \textbf{34.62}  \\
\bottomrule
\end{tabularx}
\end{table}

\section{Discussion and Findings}\label{Sec:Discussion}
Our research has effectively demonstrated the feasibility of employing Ekert's E91 as a key distribution protocol for secure image encryption. By utilizing a range of pixel-density images($64\times64$, $128\times128$ and $256\times256$ ), the encryption method consistently rendered the images unintelligible, regardless of the image's resolution. This finding underscores the potential of E91 as a robust key distribution mechanism for future secure communication applications.

Figure \ref{fig:Success_Encryption} aptly illustrates the successful encryption of an image using a seed and r-value derived from an arbitrary E91 distribution. Alice encrypts the image using the specified parameters, and Bob successfully decrypts it using the same parameters. In contrast, Figure \ref{Fig:Failed_Decryption} depicts an instance where Bob fails to decrypt the image due to a mismatch in the E91 key stream compared to Alice's. Bob attempts to decrypt the image using his keys, but the resulting chaos parameters deviate from Alice's, leading to decryption failure. Given the high sensitivity of chaotic maps to initial conditions, even minor discrepancies in the key stream can significantly impact the decryption process. Additionally, we have measured the CHSH-inequality values for maximally entangled qubits used in E91 procedure Table \ref{table:CHSH test}. Alongside Table \ref{table:CHSH test eavesdropper} demonstrates the violation of CHSH inequality when the qubits are not maximally entangled in the case of an eavesdropper. Proper transmission without any interference would result in the CHSH correlation value to be around $2\sqrt{2}$, however Table \ref{table:CHSH test eavesdropper} shows that interference will cause the correlation will deviate much further than $2\sqrt{2}$.
The performance evaluation of the proposed security scheme is summarized in Tables \ref{table:entropy}, \ref{table:Time_complexity}, and \ref{table:Diffatack}. Table \ref{table:entropy} reveals that the average information entropy for plain images is 5.91, while that of encrypted images is 7.35, approaching the maximum value of 8, indicating a high degree of randomness in the encrypted data. Table \ref{table:Time_complexity} highlights the average encryption and decryption times of the proposed scheme, which are 74.81 ns and 199.242 ns, respectively. These values demonstrate the scheme's efficiency in practical applications. Table \ref{table:Diffatack} presents the NPCR and UACI values, which measure the scheme's sensitivity to changes in the original image. An average NPCR of $99.66\%$ and a UACI of $34.62\%$ indicate that the proposed scheme is highly sensitive to the original image, making it resilient against attacks that attempt to modify the image content. Figure \ref{fig:Histogram} further corroborates the effectiveness of the proposed scheme by demonstrating the uniformity of encrypted image histograms. The absence of discernible differences from the original image's histogram reinforces the security of the encryption process. The findings presented in Figure \ref{Fig:Failed_Decryption} underscore the critical role of precise key distribution in chaotic communication. Even a minor alteration, as exemplified by flipping the first bit of the key stream, can render the decryption process unsuccessful. This observation highlights the promising potential of integrating QKD, such as E91, into chaotic communication systems to enhance security and prevent unauthorized access to sensitive data. 
 
Based on the above discussion, we observe: \textbf{(1)} The proposed scheme effectively encrypts images using a combination of chaotic communication and QKD. \textbf{(2)} The encrypted images exhibit high entropy and sensitivity to the original image.
\textbf{(3)} The scheme is efficient in terms of encryption and decryption times.\textbf{(4)} The integration of QKD enhances the security of the scheme against potential attacks.

\section{Conclusions and Future Scope}\label{Sec:Conclusion}

Our study has successfully demonstrated the feasibility and effectiveness of combining chaotic communication systems with QKD protocols to enhance the security and reliability of optical communications. The proposed scheme exhibits high entropy, efficiency, and sensitivity to original images, making it a promising candidate for practical applications in secure communication. Further research could explore the integration of E91 with other chaotic communication systems to enhance security and achieve greater robustness against potential attacks.
Further research could explore: \textbf{(1)} The integration of E91 with other chaotic communication systems to enhance security and achieve greater robustness against potential attacks. \textbf{(2)} Develop hardware implementations of the proposed scheme. \textbf{(3)} Deploy the scheme in real-world applications.
Overall, We hope that this research has made significant contributions to the advancement of secure communication technologies. The proposed scheme can potentially revolutionize how we transmit sensitive data in the digital quantum era.

\section*{Acknowledgment }
M.R.C. Mahdy acknowledges the support of NSU CTRGC Grant  2022-23 (approved by the NSU authority).

\end{document}